# *"People Are the Answer to Security":*
*Establishing a Sustainable Information Security Awareness Training (ISAT) Program in Organization*

Oyelami Julius Olusegun
Department of Information systems
University Technology Malaysia
Faculty of Computing
Skudai, Johor Bahru 81310
Jooyelami3@live. utm.my

Norafida Binti Ithnin
Department of Information systems
University Technology Malaysia
Faculty of Computing
Skudai, Johor Bahru 81310
afida@utm.my

ABSTRACT

Educating the users on the essential of information security is very vital and important to the mission of establishing a sustainable information security in any organization and institute. At the University Technology Malaysia (UTM), we have recognized the fact that, it is about time information security should no longer be a lacking factor in productivity, both information security and productivity must work together in closed proximity. We have recently implemented a broad campus information security awareness program to educate faculty member, staff, students and non-academic staff on this essential topic of information security. The program consists of training based on web, personal or individual training with a specific monthly topic, campus campaigns, guest speakers and direct presentations to specialized groups. The goal and the objective are to educate the users on the challenges that are specific to information security and to create total awareness that will change the perceptions of people thinking and ultimately their reactions when it comes to information security. In this paper, we explain how we created and implemented our information security awareness training (ISAT) program and discuss the impediment we encountered along the process. We explore different methods of deliveries such as target audiences, and probably the contents as we believe might be vital to a successful information security program. Finally, we discuss the importance and the flexibility of establishing a sustainable information security training program that could be adopted to meet current and future needs and demands while still relevant to our current users.

CATEGORIES AND SUBJECT DESCRIPTORS

*[Computer and Education]: Computer and Information Security Education*
*[Management of Computing and Information Systems]: Security and Protection*
*General Terms: Information Security, Human Factors, Management and Education*

**Keywords:** *Information Security, Awareness, End-User, Education and Training*

## I. INTRODUCTION

The essentiality and the role of information security awareness training (ISAT) should not be underestimated. ISAT program and the Information Assurance and Security Research Group of University Technology Malaysia (IASRG-UTM) with the School of Professional and Continue Education of the University Technology Malaysia (UTM-SPACE), has established and implemented a comprehensive and coherent information security awareness program to educate our users about the importance of information security (ISec). This paper will explore the creation and the establishment of the information security program, the identification of different audiences and methods of information delivery and how to define what content is vital to a successful information security program. It will also discuss how to successfully maintain a relevant and sustainable long term information security awareness program.

## II AIM AND OBJECTIVE

The goals and objectives of the ISAT program are to:

1. Change the perceptions of people's thinking and reactions when it comes to information security issues,
2. Develop a metrics as a yardstick to measure the level of knowledge of target audiences and the success of the ISAT program, and
3. To continually address the viability and importance of information security on the university premises.

## III. METHODOLOGY

UTM developed an information security awareness program for students, faculty and staff member. The program aims is to educate users and change their behavior via two main avenues as follows: (1) information security awareness training and (2) monthly activities. The methodology or planning process in achieving this will focus and consist of determining vital contents, defining audiences and choosing the correct methods of delivery.

### A. Determining the Content

In order to determine the content, the first thing we did is to evaluate the security related challenges and problems that UTM dealt with on a daily basis. We did this, based on tangible statistics, such as reports from our system users, as well as problems perceived. While having a dialogue to people about what they perceived to be our biggest security challenges and problems, we realized that some factors will always be problems and those factors will only be a problem or challenges at a specific time, and as a result of that, a new problems or challenges will always emerge. Based on these factors, we decided to incorporate flexibility with our content so we could be able to inculcate new problems or concerns as they arose. In order to accommodate this needs and the avoidance of constant revising of our material, we decided that the training component of our information security awareness training (ISAT) program would consist of topics that are static and will be evaluated on a annual basis, while the ongoing monthly activity components of our ISAT program would consist and focus of topics that were relevant at the time. Since the monthly activities focus on what is important at the time, the initial focus was to establish the list of necessary topics for the ISAT program.

After we felt it strongly that we have gotten a good idea of what should be inculcated in the ISAT program, we seek and solicited for opinions from academic managers within the UTM premises, this include our IT and desktop support team, help desk support team, server support, networking team and training managers. During this solicitation for opinions from this array of staff, we suspected and concluded that, most of the academic and non academic staff was in support and agreement with us as far as what topics should be incorporated and covered. However, some of the technical staff (Non academic) felt it strongly that we had not included enough specialized information security content to keep it more interesting. Based on this feedback, we re-evaluated the ISAT program content. In doing this, we discovered what we admitted and considered to be a more appropriate in maintaining balance between non-technical and technical information. At this juncture, our list of topics for the ISAT consisted of safety of password and security, security of workstation, emails and security of internet and physical security and protection of academic records and health data according to Buckly, (1974) and United state congress report, (1996).

In our view, we also felt it strongly that, the ISAT curriculum was a good beginning and it covered the majority of the challenges and problems UTM deals with on a daily basis, but along the process, we decided to advance further by consulting and evaluating what the information security industry would says is vital for end-user education by seeking their opinion, this become an eyes opener to another two concepts we had not considered previously that is the, social engineering which consist the integration of culture, believes and norms of the people and the principle of low or least privilege. This where not initially perceived as a major problem at MU, we now decided we would like to inculcate social engineering and the principle of least privilege to educate our users before they become a problems, although the addition of these two topics to the ISAT program for the monthly activities, we first came up with an initial list of topics with the idea and the believe that they could be adapted to meet our needs at the time. This initial list consisted of requirements for new password and digital millennium copyright according to (digital millennium copyright act) DMCA, identity theft and the university's acceptable use of information security policy.

B.  *Defining the Target Audiences*

We initially bear in mind that we would have two different audiences that is the students and the faculty/staff. While we are nurturing this idea, we quickly realized that it's not as simple as we thought as we actually have multiple or more than one audiences within the two groups and it is likely we would have more than what we have recognized so far.

C.  *Students*

The broad array of categories of students includes on-campus students living in residence halls within the university premises and off-campus students living in a self rented apartment outside the university premises. To consider these two subsets of the student population in different location will require different methods of deliveries, which will be discussed later in this paper

D.  *Staff and Faculty Member*

While most staff and faculty can be integrated into a general category for the purpose of our information security awareness program, we do recognized earlier on that many of the our faculty and staff members in administrative positions, such as deans of faculty (DOF) and head of department (HOD) belong in a category of their own. The people in these positions do not have enough time to devote in attending an enormous hours of training class or reading a long article, so we intend or have to consider their needs in a separate manners . In getting the upper level of administrators involved in the security awareness program was vital. With their signing in, we adduced that we would be more likely to obtained co-operation from the rest of their department.

## IV. METHODOLOGY FOR ISAT DELIVERY

In this particular portion of the planning process or phase was very fundamental to the success of our ISAT program. We had to consider not only the topics or the content of the program and the appropriate and adequate ways to deliver those topics, but we also need to take into consideration our different audience factions.

### A. Based On Students

In selecting our method of delivery, we have decided on a few methods of delivery that would work for all students by focusing on mass e-mail, our monthly technology newsletter articles, advertisement in the student newspaper and groups or clubs presentations. Additional methods we also planned to put in-place to reach on-campus students specifically included posters in residential and dining halls, mail-box stuffers and table-tents in all dining halls. For off-campus students, we engaged campaigns posters in the student unions, classroom buildings and frequently visited places such as the university library or computing sites, however, we had to bear in mind that exposure is not fully guaranteed as it is in the residence halls and dining places. In our observation, there are some factors that distinguish or differentiated students from faculty or staff. For instance, we can reach out to faculty and staff with in-person or personal training than their department coordinates. With students, it is much more cumbersome to coordinate training face-to-face so, we decided to concentrate and strengthen our focus on web-based training for them.

### B. Based On Faculty and Staff

For faculty and staff members, we planned and decided to use in-person and online training, campaigns poster, the monthly technology newsletter articles, payroll stuffers and targeted mass e-mails. Additionally, we also decided to make use of a concise high-level overview of the information security training to fulfill requests from administrators and people who are seeking to fix us into a preliminary scheduled meeting.

## V. IMPLEMENTATION

In the next phase, UTM began implementation on the ideas that where formulated during the planning process or phases. A comprehensive information security awareness training (ISAT) program was created that has two components: topic specific monthly activities and the general information security awareness training (ISAT) program.

### A. Our Monthly Activities

In our monthly activities UTM chooses one "hot and interesting topic" per month on which we spotlight the efforts of our information security education. The goal and objective of the monthly topic and the activities is to enhance the user's knowledge and the awareness of a particular information security challenges. Also, we hope and believed that we can get security-related information out to the campus premises in an organized manners and consistent fashion.

### B. Monthly Topic for January (Example)

The theme for January information security training (ISAT) program was "Security and Password Safety". This topic was affiliated and tied to a compulsory campus wide-range password reset campaign that was initiated. The topic was also covered in our article for information security connections newsletter ("What's the need and why change of Passwords?"). We created and mounted a poster that included instructions on passwords changing and listed password best practices to follow. We hung this poster in strategic areas where most students can observe and read, such as the computing sites, dining and residential halls. We also made it available to all departmental computer personnel support for distribution in their buildings. Furthermore, we forwarded a mass e-mail to all faculty, academic and non academic staff and students with information on the password reset campaign and general password best practices code.

### C. Monthly Topic for March (Example)

The theme for information security training program for March topic was "Cyber-Security". We invited a guest speaker from the Cybernetic Malaysia, a cyber crime task force to speak about their various on-going, current and future cyber-security efforts. We tailored and fashioned a presentation to all business and information technology (IT) classes at graduate and postgraduate level that covered issues in general security and information auditing. Finally, we created an information security awareness website that included links to and descriptions of various security sites of interest to our UTM premises and other academic milieu.

### D. Security Awareness Training

The second component of the information security awareness training (ISAT) program is our security awareness training course itself. The materials used in this course are compiled during the planning stage and process of the program. This first of this training was implemented in early January 2013.

### E. In-Person Training

The key factor of our information security awareness training (ISAT) program is currently based a one-hour, in-person training tutorial class. This class covers a wide and variety of topics, including safety of password and security, physical security and workstation and security of internet and e-mail to name a few. The ISAT is delivered without a charge to departments and students. The availability of this program was initially advertised to our computer support departmental personnel and community, who then contacted ISAG-UTM when they deem it fit to schedule their training. The course instructors are from SPACE-UTM and they do meet with each departmental support personnel prior to delivering the training program to review and preview all material and note any special circumstances or error that might exist within a particular department. It is then training would be delivered to the department. Some departments agreed to make the ISAT program mandatory while others decided to have it as an optional. This decision was left to the discretion of the department. Some group of student has also opted to take advantage of the ISAT program. These groups of students have a contact and representative of SPASE-UTM to set up the time and location for the program and classes respectively. Up to date, it has been noted and recorded that almost 900 faculty, staff and students have attended and benefited from the Information Security Awareness Training (ISAT) program.

*F. Our Online Based Training*

Another ISAT training option that we deep fit is currently under development is an online training course created by using Web-CT that would be ready to commence in the fall of 2013 precisely. This Web-CT course entails the same information that is embedded in the in-person training however; this method of delivery will allow us to expand and reach out to those users who do not have the opportunity to be served by our traditional training method. For example, we have students studying abroad, residing outside the campus, part-time students and faculty and staff members at outreach sites across the country. The online training course will allow these users to receive our information security awareness training (ISAT).

## VI. FINDINGS

We realized that our ISAT program does not address all the need require by the users, which means there is a need to adjust the program to meet their need. When adapting our ISAT program to meet our current needs, we were pleased that from the starting point we had already built in flexibility. This flexibility allowed us to make an adjustment or amends where necessary without the integrity of our ISAT program had been compromised. By being flexible and maintain the flexibility with our delivery methodology, we were able to reach out to quite a number of people, in this regard, we realized that the campus community is generally receptive to the ISAT program and they are happy to be given the opportunity to learn more about our information security awareness training.

## VII. OUR PLANS FOR IMPROVEMENT

Currently, we are hoping to work with specific academic professors especially, those who have taught computer intensive courses in all ramifications to make the Web-CT tutorial course mandatory for all students that have been enrolled or admitted into the university will also entails hypertext entry that will enable student or participant to actively add questions, comments, examples, arguments, further resource and other contribution to the text, by this all participant will be able to read and respond to the hypertext entries and create a discussion related to the lecture text. We also hope to make in-person and the online training compulsory for all staff and faculty members. In addition, we also planned to develop and enhance policies and procedures that would enable us to adequately address new information security threats or issues without having to design another information security program each time. We hope to continually identifying new delivery methods, such as working with complexes of local apartment that accommodate students to distribute fliers and mailbox stuffers. We are also looking ahead into using pre-defined communities (such as new students groups, student's residential hall, learning centers and communities) as an information dissemination avenue. Since our ISAT program is still new, the metrics to determine the level of improvement are cumbersome for us to define at the moment. For instance, we have seen an increase in reports regarding threats to information assets and computer viruses on daily bases, but we are unable to link this trend of reports to a specific cause. Are more computer viruses being circulated on the internet everyday or has our ISAT program led to the increased of report on virus infections? Acquiring statistics in this regard will allow us to measure our success of ISAT program more accurately. Finally, we also plan to continually revise the current information security awareness program to address new issues or topics, with the intention of adjusting and keeping the program relevant to our users and to the academic community as a whole.

## VIII. CONCLUSION

Information security awareness training program is required by all organization either large or small medium. Organization who see the need of protecting there valuable asset should educating the user. The users play an enormous role in information security believing and bearing in mind that, people are the key and the answers to information security that mean, people can breach information security and they can also secure it, if they lack or have the adequate and relevant information security awareness training. As many organization are envisaging new threats and challenges in information security, the information security awareness training (ISAT) program should be flexible and adjustable to meet the current challenges and that of the future by that, a sustainable information security awareness training program (ISAT) would have been established to meet the future need without jeopardizing the current. The ISAT program will also accord the users to get abreast with the knowledge of sensitive and personal data, knowledge of the organization security goal and security policies and the skills needed towards information security administration and management and to change there perceptions and reasoning when come to information security issues and also where sharing information and data exchange are required. Our flexibilities in this program, the delivery methods and the general receptiveness towards the ISAT program and the wiliness to learn more about our information security awareness training by the campus community at large has given us the impetus to further improved on the ISAT program, maintaining flexibilities and be able to reach out to more people.

## AUTHORS PROFILE

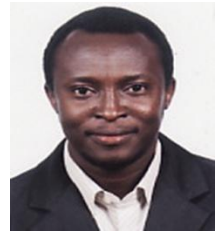**Oyelami Julius Olusegun** Is a graduate of electrical engineering technology with electrical power option from college of science and technology Ghana in collaboration with French institute of technology in 1996. Having served in engineering industry for over 10 years, in the quest for computer knowledge, he enrolled into Kursk state technical university, Russia where he obtained a B.Sc. in computer system and network engineering in 2009 and several professional certificates in IT with over ten years industrial experience in engineering. Currently is a postgraduate research student in department of information system, faculty of computing, University Technology Malaysia (UTM), and a member of information assurance and security research group (IASRG-UTM), His research interest are in information security management, social networking and information sharing and Information System. He is a professional member, association for computing machinery (ACM) and an academic member, association for information systems (AIS). He has recently extended his research interest into ICT, cloud and grid computing.

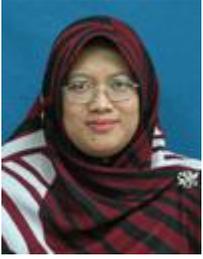 **Norafida Binti Ithnin** is currently a senior lecturer and head of department in Universiti Teknologi Malaysia (UTM), faculty of computing. She received her B.Sc degree in Computer Science (Computer Systems) form Universiti Teknologi Malaysia (UTM), Kuala Lumpur, Malaysia in 1995 and her M.Sc degree in Information Technology (Computer Science) from Universiti Kebangsaan Malaysia (UKM), Bangi, Malaysia in 1998. She bagged her PhD degree in Computation from UMIST, Manchester, United Kingdom in 2004. Currently, her main research interests are security management and graphical password.